\documentstyle[11pt,sprocl]{article}
\input{psfig}
\begin{document}
 
\def\sfb#1{\hbox{\setbox0\hbox{#1}\ignorespaces
\dimen0=\wd0
\dimen1=\ht0
\dimen2=\dp0
\advance\dimen0 2pt
\advance\dimen1 1pt
\advance\dimen2 1pt
\hbox to 0pt{\vrule depth \dimen2 height \dimen1 \hss}\ignorespaces
\hbox to 0pt{\kern1pt\copy0\hss}\ignorespaces
\hbox to 0pt{\lower \dimen2 \vbox to 0pt{\hrule width\dimen0}\hss}\ignorespaces
\hbox to 0pt{\raise \dimen1 \vbox to 0pt{\hrule width\dimen0}\hss}\ignorespaces
\kern\dimen0 \vrule depth \dimen2 height \dimen1}\ignorespaces\penalty10000}
%
%
  \vspace*{-0.5cm}
  \begin{center}
    \font\GIANT=cmr17 scaled\magstep4
	{\GIANT U\kern0.8mm N\kern0.8mm I\kern0.8mm V\kern0.8mm %
	E\kern0.8mm R\kern0.8mm S\kern0.8mm I\kern0.8mm %
	T\kern0.8mm %
	\setbox0=\hbox{A}\setbox1=\hbox{.}%
	\dimen0=\ht0 \advance \dimen0 by -\ht1%
	\makebox[0mm][l]%
	{\raisebox{\dimen0}{.\kern 0.3\wd0 .}}A\kern0.8mm
	T\kern5mm
	B\kern0.8mm O\kern0.8mm N\kern0.8mm N\kern0.8mm
	} \\[8mm]
    {\GIANT
    	P\kern0.8mm h\kern0.8mm y\kern0.8mm s\kern0.8mm i\kern0.8mm
	k\kern0.8mm a\kern0.8mm l\kern0.8mm i\kern0.8mm s\kern0.8mm
	c\kern0.8mm h\kern0.8mm e\kern0.8mm s\kern5mm
	I\kern0.8mm n\kern0.8mm s\kern0.8mm t\kern0.8mm
	i\kern0.8mm t\kern0.8mm u\kern0.8mm t\kern0.8mm
    } \\[2.1cm]
 {\Large \bf Integrable $Z_n$-Chiral Potts Model: \\  \vspace{1.5mm}
  The Missing Rapidity-Momentum Relation}\\
\vspace{1.7cm}     {\Large   G. von Gehlen }\\  \end{center}   \vspace{1cm}
{\large \bf Abstract: }
The McCoy-Roan integral representation for gaps of the
integrable $Z_n$-symme\-tric Chiral Potts quantum chain is used to calculate
the boundary of the incommensurable phase for various $n$. In the limit
$n\rightarrow\infty$ an analytic formula for this phase boundary is obtained.
The McCoy-Roan formula gives the gaps in terms of a rapidity. For the lowest
gap we conjecture the relation of this rapidity to the physical momentum
in the high-temperature limit using symmetry properties and comparing the
McCoy-Roan formula to high-temperature expansions and finite-size data.
\vspace{3cm}
\begin{figure}[h]
    \begin{minipage}{\textwidth}
\begin{raggedright}
\begin{tabular}{@{}l@{}}
Mailing address:\\Physikalisches Institut \\
 Nussallee 12 \\ 53115 Bonn, Germany   \\
 e-mail:\\ unp02f@ibm.rhrz.uni-bonn.de\\ \end{tabular}\end{raggedright}
\hfill
\parbox{5.5cm} {   {
    \vskip -1.4truecm
  \hbox to 5.5cm
  { \includegraphics{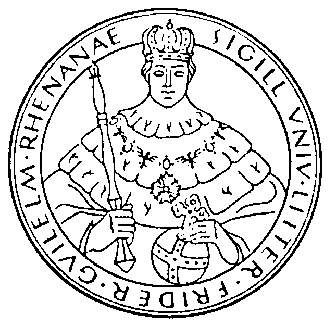}
   \hfill } \vfill  \centering } } \hfill   \begin{raggedleft}
 \begin{tabular}{@{}l@{}}
  BONN-TH-95-21    \\    hep-th/9601001\\ 
  Bonn University \\  December 1995\\       
  {\footnotesize ISSN-0172-8733}  \\        
\end{tabular}
\end{raggedleft}
    \end{minipage}
    \vspace*{5mm}
  \end{figure}
\thispagestyle{empty}
\mbox{}
\newpage
\setcounter{page}{1}
\addtolength{\oddsidemargin}{2mm}
\addtolength{\textwidth}{1cm}
\addtolength{\textheight}{-1mm}
 
\def\NPB{{\em Nucl. Phys.} B}
\def\PLB{{\em Phys. Lett.}  B}
\def\PRL{\em Phys. Rev. Lett.}
\def\JPA{{\em J.of Phys.} A: Math.Gen.}
\def\st{\scriptstyle}
\def\sst{\scriptscriptstyle}
\def\mco{\multicolumn}
\def\CPbar{\hbox{{\rm CP}\hskip-1.80em{/}}}
\newcommand{\BDM}{\begin{displaymath}} \newcommand{\EDM}{\end{displaymath}}
\newcommand{\BEQ}{\begin{equation}}  \newcommand{\EEQ}{\end{equation}}
\newcommand{\BAR}{\begin{array}}     \newcommand{\EAR}{\end{array}}
\newcommand{\BEA}{\begin{eqnarray}}  \newcommand{\EEA}{\end{eqnarray}}
\newcommand{\vpd}{\frac{\varphi}{3}} \newcommand{\vqd}{\frac{\pi-\varphi}{3}}
\newcommand{\del}{\delta}            \newcommand{\lra}{\leftrightarrow}
\newcommand{\la}{\lambda}            \newcommand{\sig}{\sigma}
\newcommand{\ep}{\epsilon}           \newcommand{\Ga}{\Gamma}
\newcommand{\al}{\alpha}             \newcommand{\thi}{\frac{1}{3}Q\pi}
\newcommand{\dq}{\bar{\partial}} \newcommand{\dt}{\partial\bar{\partial}}
\newcommand{\dd}{\partial}       \newcommand{\ds}{\partial^2\bar{\partial}^2}
\newcommand{\spi}{{\textstyle \vph=\phi=\frac{\pi}{2}}}
\newcommand{\sis}{\sin{\frac{\pi}{7}}} \newcommand{\si}{\sigma}
\newcommand{\pdr}{{\textstyle\frac{\pi}{3}}} \newcommand{\ga}{\gamma}
\newcommand{\tv}{\bar{t}\bar{v}}     \newcommand{\vtp}{\bar{v}\bar{t}_p}
\newcommand{\pn}{{\textstyle\frac{\pi}{n}}}  \newcommand{\teb}{\bar{t}}
\newcommand{\Lf}{1-2\la\cos{(\phi-\vph)}+\la^2} \newcommand{\CP}{{\cal P}}
\newcommand{\lv}{\left(\frac{1+\la}{1-\la}\right)}
\newcommand{\lf}{f(\la,\phi)}        \newcommand{\hu}{\hspace*{2mm}}
\newcommand{\sui}{\vph=\phi=\pi/2}   \newcommand{\btp}{\bar{t}_p}
\newcommand{\th}{\theta}             \newcommand{\NIF}{N\rightarrow\infty}
\newcommand{\nh}{\frac{N}{2}}        \newcommand{\vph}{\varphi}
\newcommand{\WA}{{\cal WA}_4}        \newcommand{\RIF}{R\rightarrow\infty}
\newcommand{\dme}{\delta m_1}        \newcommand{\dmz}{\delta m_2}
\newcommand{\De}{\Delta}             \newcommand{\ra}{\rightarrow}
\newcommand{\DE}{\Delta E}           \newcommand{\tp}{\tau_p}
\newcommand{\ha}{\frac{1}{2}}        \newcommand{\hb}{\frac{3}{2}}
\newcommand{\hal}{{\textstyle \frac{1}{2}}} \newcommand{\bv}{\bar{v}}
\newcommand{\OV}{\;+\;{\cal O}(\lambda^4)} \newcommand{\rsi}{\rho}
\newcommand{\pih}{\frac{\pi}{2}}     \newcommand{\AM}{(A_{m-1},A_m)}
\newcommand{\kd}{\frac{1}{3}}        \newcommand{\om}{\omega}
\newcommand{\Lb}{\left[}             \newcommand{\Rb}{\right]}
\newcommand{\lb}{\left\{}            \newcommand{\rb}{\right\}}
\newcommand{\lk}{\left(}             \newcommand{\rk}{\right)}
\newcommand{\Hn}{H^{(n)}}            \newcommand{\vm}{\vspace*{-3mm}}
\newcommand{\HH}{{\cal H}}           \newcommand{\UU}{{\cal U}}
\newcommand{\hx}{\hspace*{3mm}}      \newcommand{\hi}{\hspace*{12mm}}
\newcommand{\hs}{\hspace*{1cm}}      \newcommand{\su}{\sum_{p=1}^N}
\newcommand{\hq}{\hspace*{6mm}}      \newcommand{\ny}{\nonumber}
\newcommand{\zeile}[1]{\vskip #1 \baselineskip}
 
\title{INTEGRABLE $Z_n$-CHIRAL POTTS MODEL: \\THE MISSING
RAPIDITY-MOMENTUM RELATION}
\author{ G. VON GEHLEN \footnote{Talk given at the Satellite Meeting of STATPHYS
 19: "Statistical Models, Yang-Baxter Equation and Related Topics",
  Aug. 12-14th 1995, Nankai Institute of Mathematics, Tianjin, China}}
\address{Physikalisches Institut der Universit{\"a}t Bonn,\\
Nussallee 12, D-53115 Bonn, Germany}
\maketitle\abstracts{The McCoy-Roan integral representation for gaps of the
integrable $Z_n$-symmetric Chiral Potts quantum chain is used to calculate
the boundary of the incommensurable phase for various $n$. In the limit
$n\ra\infty$ an analytic formula for this phase boundary is obtained. The
McCoy-Roan formula gives the gaps in terms of a rapidity. For the lowest gap
we conjecture the relation of this rapidity to the physical momentum in the
high-temperature limit using symmetry properties and
comparing the McCoy-Roan formula to
high-temperature expansions and finite-size data.}
\section{Introduction: $Z_n$-symmetrical Chiral Potts Quantum Chains}
The Chiral Potts model has been introduced in 1981 by Ostlund \cite{Ost} in
order to describe commensurate-incommensurate (C-IC) phase transitions
observed in surface monolayer adsorbates, e.g. krypton on a graphite surface.
The location of a possible Lifshitz point in the phase diagram was a maior
issue. The quantum chain version has been used first by
Centen {\em et al.}\cite{CRi}. After the discovery of the novel integrability
properties of the Chiral Potts models \cite{GR,AuY}, much interest has
focussed on their mathematical aspects. Nevertheless, the spectrum of
the Chiral Potts quantum chains exhibits several unusual interesting physical
features caused by the inherent parity violation.  \par
The $Z_n$-symmetrical Chiral Potts quantum chains are defined by
\BEQ H^{(n)}=-\sum_{i=1}^{N}\sum_{k=1}^{n-1}\left\{\bar{\al}_k \,\sig_{i}^k
  +\la\,\al_k \,\Ga_{i}^k \,\Ga_{i+1}^{n-k} \right\}. \label{HN}   \EEQ
The operators $\sig_i$ and $\Ga_i$ act in a vector space $C^n$ located
at site $i$ and satisfy
\BEQ\sig_i\Ga_j=\Ga_j\sig_i\:\om^{\del_{i,j}};\hq\sig_i^n=\Ga_i^n=1;\hq
 \sig_i^+=\sig_i^{n-1};\hq \Ga_i^+=\Ga_i^{n-1}
 \label{sig}\EEQ   where $\om=e^{2\pi i/n}$.
The parameter $\la\ge 0$ plays the r\^ole of an inverse temperature (we shall
consider the ferromagnetic case only).
$\bar{\al}_k$ and $\al_k$ are complex parameters.
If we choose to represent the $\sig_i$ by diagonal matrices
$(\sig_i)_{l,m} =\om^{l-1}\del_{l,m}$ then the $\Ga_i$ become cyclic lowering
operators $(\Ga_i)_{l,m}=\del_{l,m+1}$. $N$ is the number of sites.
We use periodic boundary conditions $\Ga_{N+1}=\Ga_1$.
\par A three-parameter version of (\ref{HN}), expressing the coefficients
in terms of just two chiral angles $0\leq \vph, \phi\leq \pi$:
\BEQ \bar{\al}_k =e^{i\vph(\frac{2k}{n}-1)}/\sin{(\pi k/n)};\hi
  \al_k =e^{i\phi(\frac{2k}{n}-1)}/\sin{(\pi k/n)} \label{CCP} \EEQ
describes various cases of particular interest:
\begin{itemize}\item  For general $n$ and $\vph=\phi=0$ we recover the
Fateev-Zamolodchikov parafermionic model, for $n\!=\!2$ we
get the standard Ising quantum chain. \item If $\sui$ we get
a representation of the Onsager-Dolan-Grady-algebra : "Superintegrable
Chiral Potts-Model" \cite{GR,On}.
\item  Au-Yang {\em et al.} \cite{AuY} have shown that (\ref{HN}),(\ref{CCP})
can be derived from a two-dimensional lattice model with
(generally complex) Boltzmann weights satisfying Yang-Baxter equations
(so that $\Hn$ is integrable) if the three parameters are related by
\vm \BEQ \cos{\vph}=\la\cos{\phi}.\label{int}\EEQ \vspace{-6mm}
\item Eq.(\ref{int}) forces $\Hn$ to become non-hermitian\cite{BPA} for
$\la<1$ if we choose $|\cos{\vph}|>\la$. For (\ref{int}) with
$\cos{\vph}>\la$ the hamiltonian is parity invariant up to complex
conjugation, whereas for (\ref{int}) with $\cos{\vph}<-\la$ there is
a more complicated shifted parity symmetry \cite{APB,KYD}, see
eq.(\ref{pvp}) below.
\end{itemize}
$\Hn$ is $Z_n$-symmetrical
 and translational invariant. So, its spectrum decomposes
into sectors labeled by the $Z_n$-charge $Q=0,\ldots,n-1$ and $P$, the
momentum eigenvalue which takes the values $0\leq P=2\pi k/N <2\pi$ with
$k=0,1,\ldots,N-1$ or $-[N/2]< k \le[N/2]$.
\par  We denote the eigenvalues of $H^{(n)}$ at fixed $\la,\:\vph$ (in this
talk $\phi$ will always be determined by (\ref{int})) by $E_{Q,r}(P)$.
The index $r=0,1,\ldots$ labels the levels in each $Q, P$-sector, starting
with $r=0$ as the ground-state of the respective sector. In some part,
but not in all regions of the $\la, \vph$-plane, the ground state of
$H$ is in the $Q=P=0$-sector. By convention, we shall define gaps {\it
always} with respect to the lowest $Q=P=0$-level at the same $\la,\phi$
and $\vph$: \BEQ \DE_{Q,r}(P)=:E_{Q,r}(P)-E_{0,0}(P=0). \EEQ
Since in general parity is not conserved, a relation $E(P,\la,\vph)=
E(-P,\la,\vph)$ is expected to be valid only for special
values of $\la$ and $\vph$. Consequently, in general the minimum of
$E(P,\la,\vph)$ is not at $P=0$ as is seen e.g. in Fig.~2. Quite
generally, the spectrum of $\Hn$ seems to be a quasiparticle
spectrum \cite{DKC,GG,HG} with one quasiparticle in each $Q\neq 0$-charge
sector and one or more exceptional "Cooper-excitations" \cite{DKC}.
In the IC phase the ground state looses translational invariance,
because there the $Q=1$-quasiparticle
dispersion curve dips below the lowest $Q=P=0$-level for a certain range
of $P\neq 0$ and $\DE_{Q=1,r=0}(P)$ becomes negative.
 
\section{The McCoy-Roan gap formula}
While for the $Z_3$-superintegrable case the complete spectrum of $H^{(n)}$
has been calculated analytically \cite{APB,DKC},
only few details are known about
the spectrum in the more general integrable case (\ref{int}). In the rest of
this talk, we shall consider this integrable case. The difficulty of making
progress here comes because in this case the rapidity parameters
appearing in the Yang-Baxter-equations are restricted to higher elliptic
curves. Baxter \cite{BX} has devised an approach which avoids uniformization
of the spectral variables. Solving a functional equation\cite{APB} due to
Baxter, Bazhanov and Perk \cite{BBP}, McCoy and Roan for $Z_3$, and then
McCoy for the general $Z_n$-case \cite{MCR} have derived an analytic formula
for the low-temperature gap in the thermodynamic limit. By duality,
this is equivalent to the following expression for the lowest
$\Delta Q=1$ gap in the high-temperature regime $\la<1$:\vm  \BEA
\DE_{1,0}(\bv)\!\!&\!\!\!=\!\!\!&\!\!2(1-\la)\btp^{n/2}\pm
 2\vtp^{\;-\frac{n}{2}+1} \frac{\sqrt{(\btp^n-\!1)\lb(1+\la)^2
 -\!(1-\la)^2\btp^n\rb}}{(\om^{1/2}\vtp-1)(\om^{-1/2}\vtp-1)}
 \sin{{\textstyle \frac{\pi}{n}}} \ny \\  +n\vtp^{n/2}
 & &\hspace*{-7mm}\frac{\CP}{\pi}\int_1^{\lk\frac{1+\la}{1-\la}\rk^{2/n}}
 \hspace*{-5mm}d\teb\frac{2(\tv-\cos{\frac{\pi}{n}})}
    {(\tv)^2-2\tv\cos{\frac{\pi}{n}}-1}
 \frac{\sqrt{(\teb^n-\!1)\lb(1+\la)^2-\!(1-\la)^2\teb^n\rb}}{\teb^n-\btp^n}
 \ny \\  & & \label{MCR} \vm\EEA
Here $\CP$ in front of the integral indicates the principal value. The sign
$\pm$ in front of the second term has to be chosen $"+"$ for $0\leq\phi<\pih$
and $"-"$ for $\pih<\phi\leq\pi$. The chiral angle $\phi$ enters (\ref{MCR})
only in the combination
\BEQ \lf=\Lf\label{lf}\EEQ through $\btp^n=\lf/(1-\la)^2$.
$\bv$ is a rapidity parameter which will also be used in the form
$v =((1+\la)/(1-\la))^{1/n}\bv$. The relation of $v$ or $\bv$ to the physical
momentum $P$ will be our main concern. Only in the $Z_3$
superintegrable case it is known \cite{APB} to read $\;\;\exp{(-iP)}=
(1+\om^2 \bar{v})/(1+\om\bar{v}).$
\par We find it useful to recast (\ref{MCR}) into the following form:
\BEA\lefteqn{\DE_{1,0}(v,\la)=2\sqrt{\lf}
 \left\{1\;+2\;\Im m \frac{1}{1-\nu_p^{-1} \om^{1/2}}
\:\Im m\frac{1}{1-\la\: e^{-i(\phi-\vph)}}\right.}\ny\\& &\left.
 +\frac{1}{\pi}\CP\int_{-1}^1\;\frac{dx}{\cos{(\phi-\vph)}-x} \;2\:\Re e
\frac{1}{1-\nu^{-1}\om^{1/2}}\;\Im m\frac{1}{1-\la\:e^{i\arccos{x}}}
 \right\} \hq \label{nin}\EEA where  \BEQ
 \nu=\lk\frac{1-2\la\:x+\la^2}{1-\la^2}\rk^{\frac{1}{n}}\! v \hi\mbox{and}\hi
 \nu_p=\lk\frac{\lf}{1-\la^2}\rk^{\frac{1}{n}}\!\!v. \hx\hq  \label{taq}\EEQ
The form (\ref{nin}) has the advantage that now $v$ appears only in two
places via $\nu_p$ and $\nu$.
In the superintegrable case one has $\phi=\vph$ so that $\btp=1$ and
 in (\ref{nin}) the second term of the curly bracket vanishes. Recall that
 on the boundary of the hermitian region $\phi=0$ or $\phi=\pi$ so that
 $\cos{\vph}=\pm \la$ and we have
 $\cos{(\phi-\vph)}=\la$ and $\btp=((1+\la)/(1-\la))^{1/n}$.
\par From (\ref{nin}) we see that for $v=0,\;\infty$ the
$n$-dependence (which is due only to the factors $\om^{1/2},\;\nu$ and
$\;\nu_p$) disappears. For $v=0$ we get immediately
 \BEQ \DE_{1,0}(v=0,\la)=2\sqrt{\lf}=\left\{\begin{array}{r@{\quad:\quad}l}
  2\sqrt{1-\la^2} & \phi=0,\;\pi \\ 2(1-\la) & \phi=\pih\end{array}\right.
,\EEQ while for $v=\infty$ the principal value integral can be calculated
analytically, leading to      \BEQ
 \DE_{1,0}(v=\infty,\la)=\frac{2(1-\la^2)}{\sqrt{\lf}} =
 \left\{\begin{array}{r@{\quad:\quad}l}
  2\sqrt{1-\la^2} & \phi=0,\;\pi \\ 2(1+\la) & \phi=\pih\end{array}\right.
\label{vinf}.\EEQ
 
\section{Boundary of the region where $\DE_{Q=1}$ becomes negative.}
\par As was already mentioned at the end of Sec.1, the
C-IC phase transition arises where the ground state
looses translational invariance. This is the case where $\DE_{Q=1}(P)$,
by varying $\la$ at fixed $\vph$, as a function of $P$ acquires a second
order zero at some $P\neq 0$.
\par In the regime where parity-invariance is broken, in general
the second order zero will appear at $P\neq 0$ anyway. Therefore, even not
knowing the precise relation between $v$ and $P$, we can as well look in
(\ref{MCR}) or (\ref{nin}) for such a second order zero as a function
of $v$. In ref.\cite{MCR} McCoy and Roan have used this method
to obtain the IC phase boundary for $Z_3$. For general $Z_n$,
since the analogue of the McCoy-Roan formula for gaps with $Q\ge 2$
has not been worked out, there are only qualitative arguments
(experience from finite-size calculations) that it is the $Q=1$-gap and
not a higher-$Q$ gap which determines the C-IC phase boundary.
 
\begin{figure}[t]
\psfig{figure=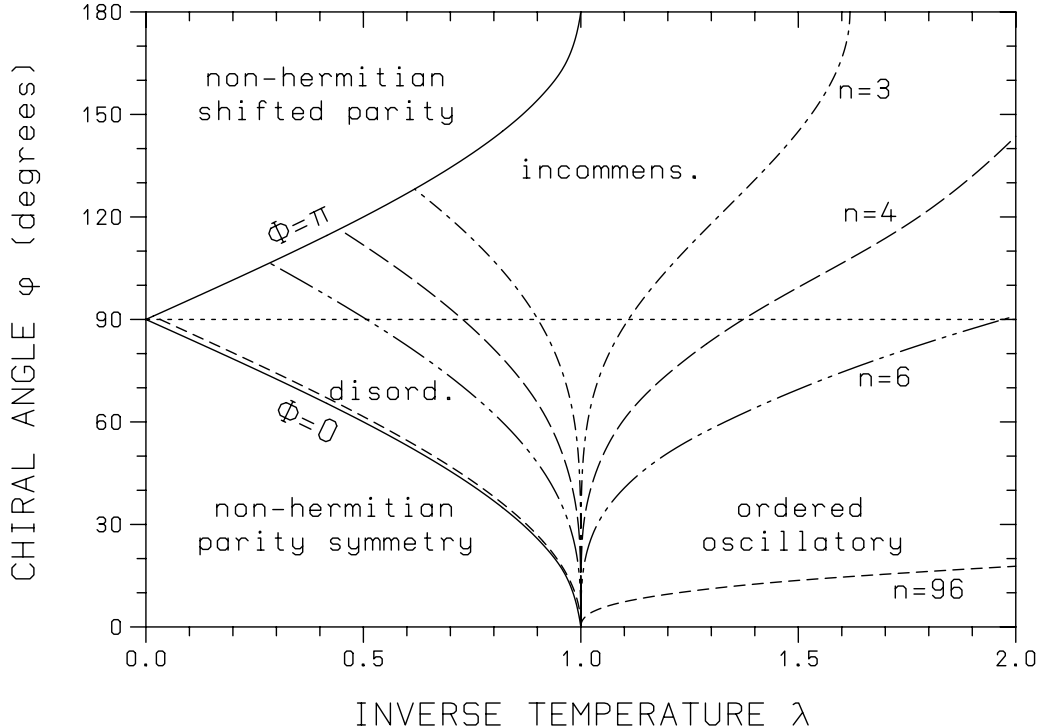,height=10cm}
\caption{Phase diagram of the integrable Chiral Potts model showing the
boundaries of the non-hermitian region $\phi=0,\;\pi$ (full curves). The
dashed-dotted curves give the boundaries of the IC phase for the $Z_n$-models
with $n=3,\;4,\;6,\;96$. \label{fig:pha}}     \end{figure}
 
\par For the $\DE_{Q=1}$-gap we
did the calculation for many values of $n$ for various $\vph$ and
obtain the curves shown in Fig. 1. The low- and high-temperature IC-phase
boundaries are related by duality: $\la\lra\la^{-1}$ together with
$\vph\lra\phi$.
\par  Looking at our numerical results for very large $n$
($n=10^3\ldots 10^5$) we found that in the limit $\;\;\sqrt{n}\sin{(\vph/2)}
\ra\infty$ the phase boundary curve $\la_c(\vph)$ and the corresponding
value of $v$ at the second order zero, $v_c$, assume the simple form
(here we give the values $\la_c$ for the low-temperature branch):
\vspace*{-3mm}  \BEQ\la_c=\frac{2n}{\pi}\mbox{sin}^2(\vph/2) +\ldots;\hi
 v_c = 1-\frac{\pi}{n}\cot{(\vph/2)}+\ldots. \label{ays} \EEQ
For $\vph\ll 1$ the IC phase forms a wedge around $\la=1$
described by $\;\la_c-1=a_n\vph^{2n/(n-2)}+\ldots \;$ where
$a_3=0.0185(2);\; a_4=0.0883(3);\; a_5=0.153(1);\; a_6=0.204(2).$
Notice that for small $n$ this is very narrow.
\par Once knowing the result (\ref{ays}) it is not difficult to prove it
analytically from (\ref{nin}) expanding $\la_c^{-1}$ and $v_c-1$ in powers
of $q=\pi/n$. Recall that the $n$-dependence comes in (\ref{nin}) only
through  $\om^{1/2},\;\nu$ and $\;\nu_p.$ Two principal value integrals
appear, which can be calculated explicitly.
 
\begin{figure}[t]
\psfig{figure=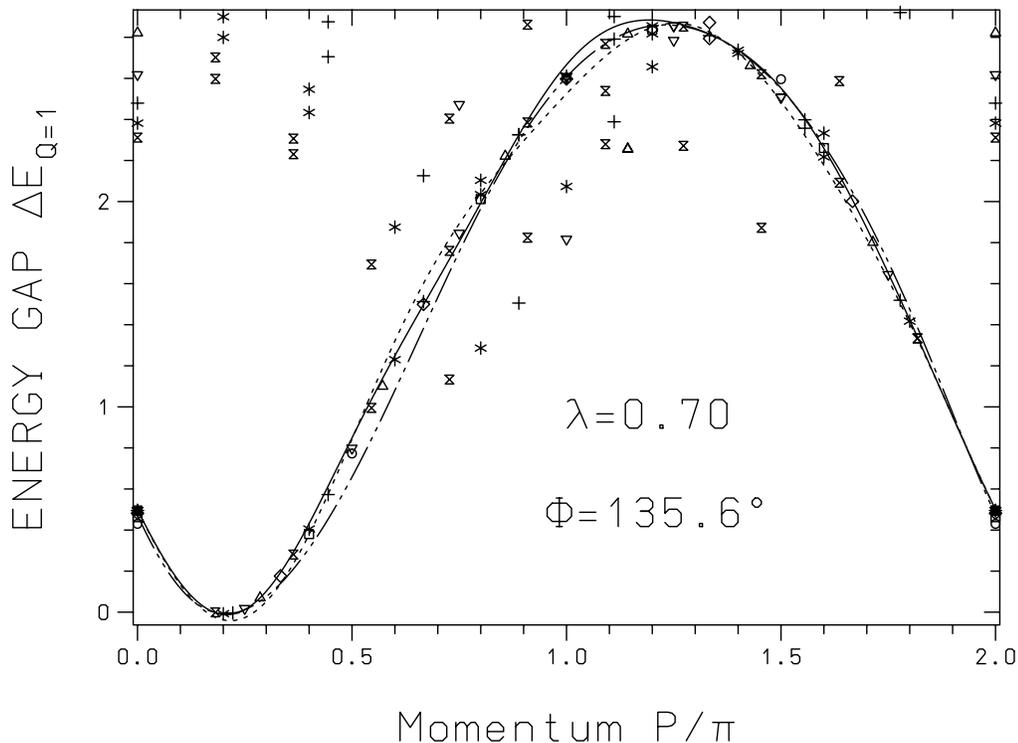,height=10cm}
\caption{Energy-momentum plot of low-lying levels in the $Q=1$-sector of the
$Z_3$-integrable Chiral Potts model for $\la=0.7$ and $\vph=120^\circ$ (which
is close to the C-IC-phase boundary since $\DE_{Q=1}$ is just getting negative
at $P\approx 0.2\pi$). Dashed-dots curve: McCoy-Roan formula using our
rapidity-momentum relation (16). Full and dotted curves:
expansion to 4th and 3rd order in $\la$, respectively. Symbols:
finite-size data for up to $N=11$ sites.\label{fig:dis}}     \end{figure}
 
\section{Determination of the Rapidity-Momentum Relation}
In view of the difficulties with high genus Riemann surfaces, probably it
will take some time until someone will come up with an analytic derivation
of the dependence $P(v)$ for the gap $\DE_{Q=1}$ at general $n,\;\la$
and $\vph$. Therefore we find it worthwhile to do what is possible right away,
which is to combine numerical and symmetry information for deducing the
main behaviour of $P(v)$:
\par 1) Numerical diagonalization of $H^{(n)}$ for up to $N=3\ldots 12$
sites is easy and leads to a quite precise determination of the gap
$\DE_{Q=1,r=0}(P)$ over most part of the region $0\le\la\le 0.9$ and
$0\le\vph<120^\circ$.
A third order high-temperature expansion (which we performed up to
$n=8$) leads to almost the same values even up to
$\la\approx 0.7$, only at larger $\la$ there are problems for
$P\le \pih$, see the Figures for $Z_3$ given in ref. \cite{HG}.\\
All these results show that $\DE_{Q=1,r=0}(P)$ is smooth
with just one maximum and one minimum in the first Brillouin zone of $P$.
Also (\ref{nin}) for fixed $n,\;\la,\;\vph$ as a function of $v$ has only one
single maximum and minimum in the whole range $-\infty<v<+\infty$. Therefore
numerically it is easy to determine the mapping $P(v)$ which brings
both curves into coincidence.
\par 2) Applying this method to the superintegrable line, after some
educated guesswork we find that the $Z_3$ formula quoted in the text before
eq.(\ref{taq}), should be generalized to $Z_n$ as:
\BEQ e^{-iP}\!=\frac{1-\om^{1/2}\bar{v}}{1-\om^{-1/2}\bar{v}}\hq\mbox{or}\hq
\bar{v}=-\frac{\sin{\hal P}}{\sin{\hal(P-P_\infty)}}
\hu\mbox{with}\hu P_\infty=\frac{2\pi(n-1)}{n}. \label{svn} \EEQ
\par 3) We want to summarize also other numerical results in a compact
empirical formula (which for $\vph\ra\pih$ has to reduce to (\ref{svn})). For
this purpose we use the fact that at the boundaries of the hermitian region
$\phi=0,\:\pi$ there are special reflection symmetries \cite{KYD}, which
follow from eq.(B.18) of Albertini {\em et al.}\cite{APB}, and which
determine $P(v)$ at $v=\pm 1$:\\
$\bullet\;\;$ At the lower boundary $\phi=0$ (see Fig.1) $\Hn$ is
parity invariant. In (\ref{nin}) this is feature shows up as symmetry
\footnote{Observe that the symmetry is in $v$ and {\em not} in $\bv$}
under $v\lra v^{-1}$ which appears just at $\phi=0$. So {\em at $\phi=0$},
comparing also the numerical data, for $0\le\la<1$ we
get \BEQ P(v)=2\pi-P(v^{-1});\hx\mbox{so that}\hx
P(v=1)=\pi;\hx P(v=-1)=0. \label{pvn} \EEQ
$\bullet\;\;$
For $\phi=\pi$ the spectra have a charge-sector-dependent shifted
parity sym\-metry \cite{APB,KYD}, which again corresponds to a symmetry of
(\ref{nin}) against $v\lra v^{-1}$. Using (B.18) of ref. \cite{APB}
we find at $\phi=\pi$ for $0\le\la<1$
\BEA P(v)&\!=\!&2\pi-P(v^{-1})-4\pi/n\hq\ny\\ \hspace{-12mm}
\mbox{so that}&\hspace{1mm}&P(v=1)=(n-2)\pi/n;\hq P(v=-1)=2(n-1)\pi/n.
 \hspace{9mm} \label{pvp} \EEA
\par By making the following ansatz for $\nu_p$ ($\nu_p$ is proportional
to $v$, see (\ref{taq})), we fulfil all constraints
(\ref{pvn}),~(\ref{pvp}) and for $\vph=\phi=\pih$ obtain (\ref{svn}):
\BEQ \nu_p =-\frac{\sin{\hal(P-P_0)}}{\sin{\hal(P-P_\infty)}}\hx
\mbox{with}\hx P_0=-2(\phi-\vph)/n;\hq P_\infty=2\pi
-2(\phi+\vph)/n.  \label{svm} \EEQ
\par We have calculated various dispersion curves using (\ref{nin}) with
(\ref{svm}) and compared these to finite-size- and perturbative data for
the $Z_3$ to $Z_7$-models. Because of space limitations, we show only Fig. 2
which gives a non-trivial example. For $0\le\la<0.4$ the agreement is so good
that we conjecture (\ref{svm}) to be exact in the limit
$\la\ra 0$. A disagreement arises if we move towards the parafermionic
corner $\la\ra 1$, $\phi\ra 0$. Here the momentum
interval corresponding to positive $v$ expands such that finally there is
only a small interval around $P=2\pi$ left for the whole negative region
$-\infty<v<0$. Put differently, both $P_\infty$ and $P_0$ seem to converge
to $P=0$ as we move towards $\la=1$ on the curve $\phi=0$. The ansatz
(\ref{svm}) does not yet take care of this behaviour. For the $Z_3$-case and
$0.95<\la<0.999,\;\phi=0$ a good fit can be obtained neglecting the negative
$v$-region altogether using $v = \mid\tan{\frac{P}{4}}\mid^{2/3}$. There
$\DE_{Q=1,r=0}(P)$ becomes $\la$-independent with a peak value of
$\DE_{Q=1}=6$ at $P=\pi$ or $v=1$.
 
\section{Conclusions}
Comparing the McCoy-Roan formula for the lowest gap of the integrable
$Z_n$-Chiral Potts model to finite-size and perturbative data and using the
special symmetry properties at the
boundary of the hermitian regime, we obtain information about the hitherto
unknown relation between the rapidity para\-meter and the physical momentum
{\em off} the superintegrable line.
The numerical agreement suggests that the conjectured formula (\ref{svm})
may be exact in the high-temperature limit. We also find an analytic formula
for the IC-phase boundary curve of the integrable $Z_n$-Chiral Potts model
valid for large $\sqrt{n}\sin{(\vph/2)}$. For $n\ra\infty$
the whole superintegrable line will be in the IC-phase.
 
\section*{Acknowledgements}
The author thanks Barry M.McCoy, Michael Baake, Fabian Essler,
Andreas Honecker and Karim Yildirim for fruitful discussions.

\end{document}